\documentclass[twoside,10pt,a4paper]{newFNLstyle}
\usepackage{graphics}
\usepackage{cite}

\def\w0{\omega_0}

\begin{document}

\volnumpagesyear{0}{0}{000-000}{2005}
\dates{15 January 2005}{23 April 2005}{6 May 2005}

\title{ROLE OF THE COLORED NOISE IN SPATIO--TEMPORAL BEHAVIOR OF TWO COMPETING SPECIES}

\authorstwo{D. Valenti$^{\circ}$, A. Fiasconaro and B. Spagnolo}
\affiliationtwo{Dipartimento di Fisica e Tecnologie Relative and
INFM, Group of Interdisciplinary Physics\footnote {Electronic
address: http://gip.dft.unipa.it.}} \mailingtwo{Universit\`a di
Palermo, Viale delle Scienze pad. 18, I-90128 Palermo, Italy \\
$^{\circ}$valentid@gip.dft.unipa.it}


\maketitle

\markboth{D. Valenti, A. Fiasconaro and B. Spagnolo}{Pattern
Formation in the Presence of Colored Noise}

\pagestyle{myheadings}


\vskip -0.6cm

\centerline{\footnotesize Communicated by Werner Ebeling and
Bernardo Spagnolo}


\vskip 0.4cm

\begin{abstract}
We study the spatial distributions of two randomly interacting
species, in the presence of an external multiplicative colored
noise. The dynamics of the ecosystem is described by a coupled map
lattice model. We find a nonmonotonic behavior in the formation of
large scale spatial correlations as a function of the multiplicative
colored noise intensity. This behavior is shifted towards higher
values of the noise intensity for increasing correlation time of the
noise.
\end{abstract}

\section{Introduction}

Recently there is an increasing research activity on the role of
noise in spatially extended systems \cite{Zaikin,Sancho,Katja}.
One of the challenging problems to be understood is the formation
of large scale spatial correlations. In this work we present a
theoretical investigation on the role played by the colored noise
in the transient dynamics of the spatial distributions of two
competing species. The colored noise in spatially extended systems
has been investigated in the past \cite{Sancho} (and references
cited there), \cite{Garcia-Sancho}. Aim of the work is to present
a stochastic model with a realistic noise and to suggest a
possibility of application to biological systems, where the
presence of fluctuations, such as random variability of
temperature, can modify strongly the dynamics of an
ecosystem~\cite{Zimmer,Bjornstad,Ciuchi}. The theoretical results
in fact may be compared with the experimental data obtained for
the biomass of species measured in the Sicily
Channel~\cite{Garcia}. Our investigation could contribute (i) to
select environmental forcings which affect the dynamics of
biological systems~\cite{Bjornstad,Bjornstad2,Alley}, and (ii) to
implement physical models for the interpretation of spatial
distributions of marine species.

\section{The Model}

To study the colored noise effect on the formation of large scale
spatial correlations of two species densities, we consider a coupled
map lattice model~\cite{Kaneko,Valenti}

\vskip -0.2cm

\begin{eqnarray}
x_{i,j}^{n+1}&=&\gamma x_{i,j}^n (1-x_{i,j}^n-\beta^n y_{i,j}^n)+
x_{i,j}^n (\zeta_{i,j}^n)_x + D\sum_\delta
(x_{\delta}^n-x_{i,j}^n),
\label{Lotka_eq_1}\\
y_{i,j}^{n+1}&=&\gamma y_{i,j}^n (1-y_{i,j}^n-\beta^n x_{i,j}^n)+
y_{i,j}^n (\zeta_{i,j}^n)_y + D\sum_\delta
(y_{\delta}^n-y_{i,j}^n), \label{two} \label{Lotka_eq_2}
\end{eqnarray}

\vskip -0.2cm
\noindent where $x^n_{i,j}$ and $y^n_{i,j}$ are the densities of
two species in the site ($i,j$) at the time step $n$, $\gamma$ is
proportional to the growth rate, $D$ is the diffusion coefficient,
$\sum_\delta$ is the sum over the four nearest neighbors, and
$\beta^n$ is the interaction parameter. Here $(\zeta_{i,j}^n)_x$
and $(\zeta_{i,j}^n)_y$ are Ornstein--Uhlenbeck colored noises at
each site ($i,j$), and correspond to the value of the stochastic
continuous process $\zeta_l(t)$ given by the equation

\vskip -0.2cm
\begin{equation}
\frac{d\zeta_l(t)}{dt}=-\frac{1}{\tau_c}\zeta_l(t) +
\frac{1}{\tau_c} \xi_l(t), \qquad (l=x,y) \label{colored_noise}
\end{equation}
taken at the time step \emph{n}. Here $\xi_i(t)$ $(i=x,y)$ are
Gaussian white noises within the Ito scheme with zero mean and
correlation function $\langle \xi_l(t)\xi_k(t')\rangle = \sigma
\delta(t-t')\delta_{lk}$. The correlation function of the
processes of Eq.~($\ref{colored_noise}$) is
\begin{equation}
\langle \zeta_l(t)\zeta_k(t')\rangle = \frac{\sigma}{2 \tau_c}
e^{-|t-t'|/\tau_c} \delta_{lk}
 \label{correlation function}
\end{equation}
and gives $\sigma \delta(t-t')\delta_{lk}$ in the limit $\tau_c
\rightarrow 0$. In absence of noise, for $\beta < 1$ both species
survives in a coexistence regime, while for $\beta > 1$ one of the
two species vanishes after a certain time and an exclusion regime
takes places. To describe the noisy interaction between species
and the environment we consider an Ito stochastic differential
equation for the parameter $\beta$ with a periodical driving
force~\cite{Valenti,Spagnolo,Valenti1}

\begin{equation}
\frac{d\beta(t)}{dt} = -\frac{dU(\beta)}{d\beta}+ a cos(\omega_0
t) + \xi_{\beta}(t)
\label{beta_eq}.
\end{equation}
Here $U(\beta)$ describes the two stable states of the ecological
system (coexistence and exclusion)

\begin{equation}
U(\beta) = h(\beta-(1+\rho))^4/\eta^4-2h(\beta-(1+\rho))^2/\eta^2.
\label{U(beta)}
\end{equation}
The values of the parameters $h$, $\rho$ and $\eta$ depend on
intrinsic biological characteristics of the species, as stability
of their life cycles respect to external perturbations and ability
of one species to compensate the increase or decrease of the other
species. The periodical driving mimics the climatic temperature
oscillations, and $\xi_\beta(t)$ is a white Gaussian noise with
$<\xi_\beta(t)>=0$ and
$<\xi_\beta(t)\xi_\beta(t')>=\sigma_\beta\thinspace \delta(t-t')$.
The interaction parameter $\beta^n$ of Eqs.~(\ref{Lotka_eq_1}),
(\ref{Lotka_eq_2}) corresponds to the value of the stochastic
process $\beta(t)$ of Eq.~(\ref{beta_eq}) taken at the time step
\emph{n}. In the absence of the additive noise, $\beta(t)$
oscillates periodically and its values always remain below $\beta
= 1$, giving rise to a coexistence regime. We fix the additive
noise intensity at the value $\sigma_\beta=2.65 \cdot 10^{-3}$
corresponding to a competition regime with $\beta$ periodically
switching from coexistence to exclusion regions. This produces the
appearance of stochastic resonance (SR) effect in the dynamics of
interaction parameter $\beta$ and noise-induced anticorrelated
periodic oscillations in the time evolution of the two species
densities~\cite{Valenti,Valenti1}. We analyze the spatial
distribution of the ecosystem in this SR dynamical regime, by
varying both the intensities of multiplicative colored noise in
Eqs. (1) and (2).

\subsection{Spatial distributions}
 By using a square spatial lattice with $N = 100 \times 100$ sites,
and species densities vanishing on the grid borders, we obtain the
spatio-temporal distributions of the two species for uniform initial
condition and for three values of the correlation time $\tau_c$,
namely $\tau_c = 10^{-2}, 10, 10^2$. We find that for the lowest
value of the correlation time ($\tau_c = 10^{-2}$), the spatial
distribution is almost the same as that found for multiplicative
white noise \cite{Valenti}.
\begin{figure}[htbp]
\centering{\resizebox{13cm}{!}{\includegraphics{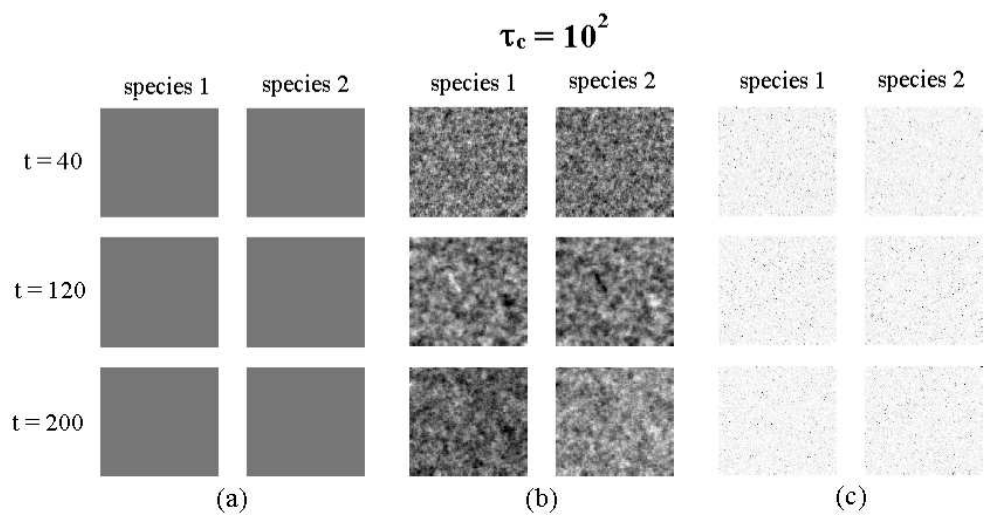}}}
\caption{Spatial distributions with $\tau_c = 10^2$ at different
times for (a) $\sigma= 10^{-12}$, (b) $\sigma= 10^{-4}$, (c)
$\sigma= 10^5$. The values of the parameter set are:
$\sigma_\beta=2.65 \cdot 10^{-3}$, $\gamma = 2$, $D = 0.05$, $a =
1.5 \cdot 10^{-1}$, $\omega_0/(2\pi) = 10^{-2}$, $\beta(0)=0.94$,
$N = 100 \times 100$, and $x^0_{i,j}=y^0_{i,j}=0.5$ at all sites
$(i,j)$.} \label{fig1}
\end{figure}
In Fig.~\ref{fig1} the spatial distributions of the two species
are obtained with correlation time $\tau_c = 10^2$ and for $\sigma
= 10^{-12}, 10^{-4}, 10^{5}$. We see that for very low noise
intensity (Fig.~\ref{fig1}(a)) the spatial distributions of the
species are uniform as at $t = 0$. For higher noise intensities
(see Fig.~\ref{fig1}(b)) an anticorrelation between the two
species is observed: the two species tend to occupy different
positions. Further increase of the noise causes the
anticorrelation to disappear and the two species densities become
uncorrelated (see Fig.~\ref{fig1}(c)). We observe that
anticorrelated behavior is stronger for increasing $\tau_c$. In
particular by comparing the distributions obtained with different
values of correlation time $\tau_c$ we note that for $\tau_c =
10^{2}$ the two species densities are strongly anticorrelated.
This is expressed by the presence, in the same spatial region, of
a high density (dark zones) of species 1 (species 2) corresponding
to a low density (light zones) of species 2 (species 1) (see
Figs.~\ref{fig1}(b)).

\subsection{Correlation between species}

\begin{figure}[htbp]
\centering{\resizebox{12cm}{!}{\includegraphics{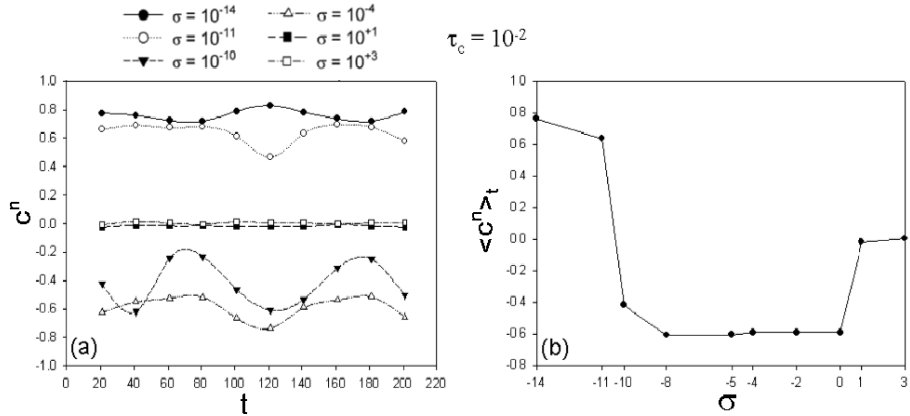}}}
\caption{(a) Correlation coefficient $c^n$ as a function of time,
for $\tau_c = 10^{-2}$ and for different values of the
multiplicative noise intensity $\sigma$, namely $\sigma = 10^{-14},
10^{-11}, 10^{-10}, 10^{-4}, 10^{+1}, 10^{+3}$; (b) time average
correlation coefficient $<c^n>_t$ as a function of the
multiplicative noise intensity.} \label{fig2}
\end{figure}

To analyze the role of the colored noise on the correlation
between the two species, we calculate for different values of the
correlation time $\tau_c$, at the time step $n$, the correlation
coefficient $c^n$ defined on the lattice as
\begin{equation}
c^n=\frac{cov^n_{xy}}{s^n_x s^n_y} \label{correlation}
\end{equation}
with
\begin{equation}
cov^n_{xy}=\frac{\sum_{i,j}(x^n_{i,j}-\bar{x}^n)(y^n_{i,j}-\bar{y}^n)}{N}
\label{covariance},
\end{equation}
where $\bar{x}^n$, $s^n_x$, $\bar{y}^n$, $s^n_y$ are the mean value
and the root mean square respectively of species 1 and species 2,
obtained over the whole spatial grid at the time step $n$,
$cov^n_{xy}$ is the corresponding covariance and $N = 100\times 100$
is the number of sites of the considered lattice. The behavior of
the correlation coefficient $c^n$ as a function of time has been
obtained for $\tau_c = 10^{-2}, 10, 10^2$, and for different levels
of the multiplicative colored noise. For $\sigma = 0$ the species
are strongly correlated and $c^n$ is constant. In Fig.~\ref{fig2} we
report the behavior of $c^n$ for $\tau_c = 10^{-2}$.
For low levels of the multiplicative colored noise ($\sigma =
10^{-14}$, $\sigma = 10^{-11}$), the values of $c^n$ oscillate
between $0.5$ and $0.8$, showing a stronger correlation between
species for lower values of $\sigma$. For intermediate values of
noise intensity, namely for $\sigma = 10^{-10}, 10^{-4}$ the
spatial correlation coefficient $c_n$ oscillates between $-0.6$
and $-0.2$ at the frequency of the periodical forcing, showing
strong anticorrelation between the two species. By increasing the
noise intensity, the anticorrelation is reduced and finally it
disappears ($\sigma = 10^{+1},\sigma = 10^{+3}$), that is the
species are totally uncorrelated.  A nonmonotonic behavior of the
correlation coefficient as a function of the multiplicative
intensity appears clearly. This behavior is more evident by
considering the time average correlation coefficient $<c^n>_t$ as
a function of the multiplicative noise intensity (see
Fig.~\ref{fig2}(b)). The minimum in Fig.~\ref{fig2}(b) corresponds
to the anticorrelated oscillations shown in Fig.~\ref{fig2}(a).
Moreover this behavior is connected with the anticorrelated
oscillations present in the time evolution of two competing
species in each point of our spatial grid \cite{Valenti1}. By
comparing the results shown in Fig.~2 with the white noise case
\cite{Valenti} we observe that the effect of a small correlation
time $\tau_c$ is to produce an enlargement of the minimum of
$<c^n>_t$ towards higher values of the multiplicative noise
intensity. The ecosystem remains in his anticorrelated regime for
a longer time due to the memory of the colored noise.
\begin{figure}[htbp]
\centering{\resizebox{12cm}{!}{\includegraphics{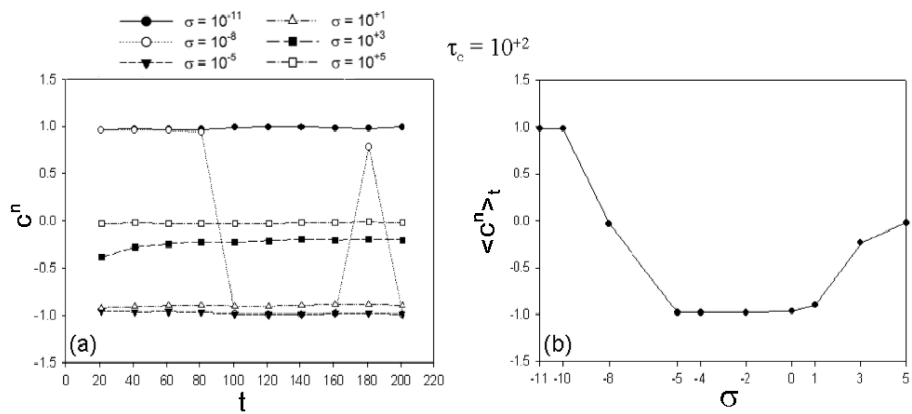}}}
\caption{(a) Correlation coefficient $c^n$ as a function of time,
for $\tau_c = 10^2$ and for different values of the multiplicative
noise intensity $\sigma$, namely $\sigma = 10^{-11}, 10^{-8},
10^{-5}, 10^{+1}, 10^{+3}, 10^{+5}$; (b) time average correlation
coefficient $<c^n>_t$ as a function of the multiplicative noise
intensity.} \label{fig3}
\end{figure}
The two noise sources have different roles on the dynamics of our
ecosystem. Specifically the additive noise affects directly the
dynamical regime, while the multiplicative noise breaks the
symmetric dynamical behavior of the two species by producing a
coherent response of the system~\cite{Valenti,Valenti1}, which is
responsible for the appearance of anticorrelation in the spatial
distributions of the two species. In Fig.~\ref{fig3} the behavior
of $c^n$ and $<c^n>_t$ is reported for $\tau_c = 10^2$. As in the
previous case we observe a nonmonotonic behavior of $c^n$ as a
function of the multiplicative noise intensity. For low noise
intensities ($\sigma = 10^{-11}$, $\sigma = 10^{-10}$), the
species are strongly correlated and $c^n$ is almost constant. For
$\sigma = 10^{-8}$ strong oscillations between $+1.0$ and $-1.0$
appear. This effect is due to the colored noise, in fact the
ecosystem remains periodically in the correlated or anticorrelated
regime for a time interval of the order of the correlation time of
the noise. For greater noise intensity ($\sigma = 10^{-5}, 10^1$),
the anticorrelation regime takes place ($c^n \approx -1.0$). By
increasing the noise intensity ($\sigma = 10^{+3}, 10^{+5}$),
$c^n$ goes through a nonmonotonic behavior, taking on values
around zero (uncorrelated regime). Comparing Fig.~\ref{fig2} and
Fig.~\ref{fig3} we observe that the presence of a bigger
correlation time $\tau_c$ determines a stronger anticorrelation in
the two species distributions. The minimum of the time average
correlation time $<c^n>_t$ is deeper and it is shifted towards
higher values of the multiplicative noise intensity for greater
values of correlation time $\tau_c$. A highly correlated noise
therefore produces two effects: (i) the system is pushed towards a
strongly anticorrelated regime ($c^n\approx -1.0$); (ii) this
regime takes place for higher values of multiplicative noise
intensity. The colored noise produces therefore a persistence of
the dynamical regime experienced by the ecosystem.

\section{Conclusions}

In this work we analyze the effects of the colored noise in the
spatial distributions of two competing species. By using a
discrete time evolution model, which is the discrete version of
the Lotka--Volterra equations with diffusive terms, in the
presence of a multiplicative colored noise, with a random
interaction parameter, we analyze the spatio-temporal behavior of
the two species. We find that the colored noise determines an
enhancement of the effects observed in the presence of white
noise~\cite{Valenti}, that is (i) formation of large scale spatial
correlations with the same periodicity of the deterministic force
and (ii) anticorrelation in the spatial distributions of the two
species. Finally we observe that (i) the intensity of the
anticorrelation increases as a function of $\tau_c$, (ii) the
system remains in the anticorrelated region for a bigger range of
multiplicative noise values as $\tau_c$ increases. The noise
spectrum of a real physical system is characterized by a cut-off
(correlated noise). Therefore data from real ecosystems, whose
dynamics is strongly affected by fluctuations and by environmental
physical variables, can be modeled using sources of colored noise.

\section{Acknowledgements}

This work has been supported by INTAS Grant 01-450, INFM and MIUR.

\end{document}